# The Distribution of Vaccine among Countries in the Case of a Pandemic

**Béla Vizvári[1]**, **Gergely Kovács[2]**

[1] Dept. of Industrial Engineering, Eastern Mediterranean University, bela.vizvari@emu.edu.tr

[2] Edutus University, Tatabánya, Hungary, kovacs.gergely@edutus.hu

## Abstract

**Purpose:** Only few companies were able to produce vaccine again COVID-19. Thus, one producer supplied it to many countries. The distribution was not effective. Some countries overstocked the vaccine while other countries were not able to buy enough. The purpose of the present paper is to provide with a frame such that one producer distributes the vaccine to a set of countries in a way that the shortage is minimized.

**Methodology:** The consumption of the countries are approximated by regression functions taking into account the saturation of the process. The distribution of the vaccine is determined by MIP models of operations research.

**Findings:** Effective distribution of vaccine can be obtained for even a large number of countries. Both the level of the shortage and the number of the consecutive shortage days in a country can be controlled.

**Practical implications:** A group of countries can act as a single partner of a pharmaceutical company. They can get a steady supply. The company gets a well-organized delivery plan.

**Social implications:** More people can be saved because of the steady supply of the vaccine.

**Originality:** The paper develops a new concept for the fair distribution of vaccines between countries. This concept and the method derived from it can be applied in the event of future pandemics on a global scale.

## 1. Introduction

The COVID-19 pandemic was similar to the Spanish flu pandemic of 1918-1919. The latter one covered the whole world as COVID-19 did. An important difference is that information technology has been developed meanwhile. Thus, there are detailed data on the course of the pandemic in all countries of the world (database 2023). The database contains daily data on the number of cases, and the number of vaccinations in each country.

An important feature of a pandemic is saturation. First, it spreads slowly. The speed is increasing. The number of the daily cases is almost constant in the middle of the epidemic. Then, the number of the daily new cases is decreasing and becomes small. If the disease has the property that a person can get it only once, then the size of the population is an upper bound to the total number of cases. It is not completely true for COVID-19. However, the size of the population was still a rough upper bound on the total number of cases in each country. What is important here is the existence of the upper bound and not its particular value. The upper bound causes the saturation. A saturation process can be approximated by the sigmoid function and also by the arcus tangent function. The latter one also considered frequently as one of the sigmoid functions. Good quality regression can be obtained for the data of COVID-19 with correlation between the original data and the regression function with correlation higher than 0.99 (Pınarbaşı, Vizvári, 2024). This fact is the basis of the results of this paper.

Another important lesson of COVID-19 is that there was no vaccine at all at the beginning of the pandemic. Even later only few pharmacological companies were able to develop and produce vaccine. It is well-known that the production of vaccine is sensitive. If the production is stopped, then it is difficult and time consuming to start the process again. Thus, both the limited production capacity and the technical stops may create shortage.

The vaccination of the individuals is the task of the health care system. The procurement of the vaccines, the distribution of the purchased quantity among private or state-owned health care units, and the determination of the prioritized groups of the population is the task of the state authority.

Turning to the scenario of a future pandemic, it is supposed that producer cannot be found in every country. Thus, a production unit will deliver vaccine to several countries. Countries can compete with each other or cooperate. The latter means that they act as a single buyer and allocate among themselves the quantity made available to them by the manufacturer in a manner determined by them. The main topic of this paper is the method how the allocation can be determined.

**2. The nature of the demand function**

The basic properties demand function are discussed in a static environment in this section. The COVID-19 pandemic is used as an example as the saturation property implies that the behavior of the demand must be similar in the case of other pandemics as well. How this structure can be applied in a dynamic and partially unknown environment is analyzed in section XX.

The demand function of a country is the number of used vaccines as the function of time. It is what can be approximated by a saturation, *i.e.* sigmoid or arcus tangent, function. Figure 1 shows an example. It is on the total vaccines used in Mexico. The value of the correlation of the original and the fitted function is 0.99475. The regression function is

$$733420000 \frac{\arctan(5.32 \times 10^{-9}(DAY-322.96))}{3.14}. \qquad (1)$$

The first day of vaccination is 2020.12.24. The last day on the figure is Day 653 which is 2022.10.07.

The demand function of a country depends on several factors. The size of the population is important. The willingness to be vaccinated depends on the local culture and the danger caused by the pandemic. There are random effects as well. A very low percentage of the population is sensitive for the vaccine in negative sense. If such a case gets a lot of publicity, then the willingness can be reduced for a while. Moreover, counter-propaganda against vaccination always exists. Its impact depends on the awareness of society and can also be considered accidental.

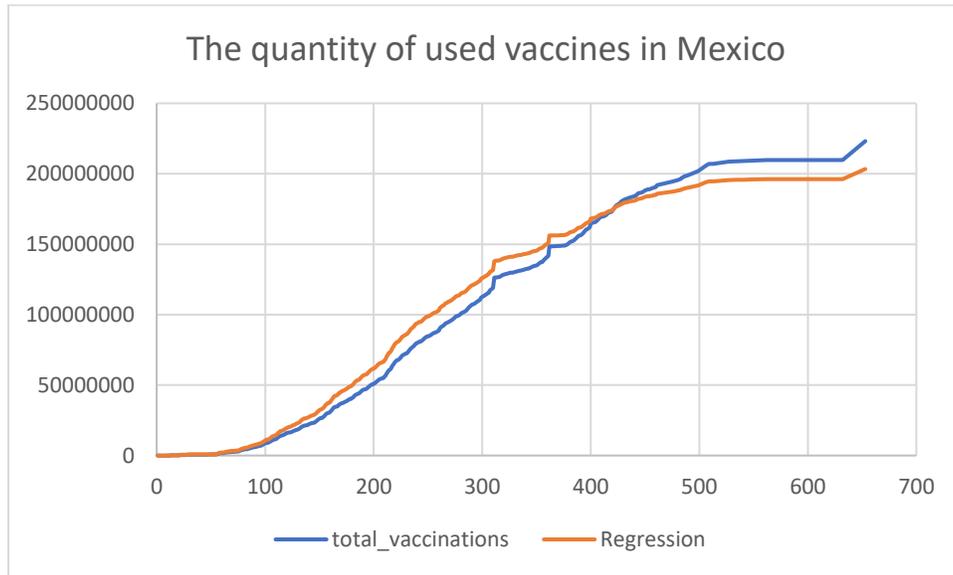

Figure 1. Regression function of the quantity of the used vaccines in Mexico.

The demand function is the sum of functions of type (1) in the multi-country case. Figure 2 shows the total demand of five countries between 2020.12.27 and 2023.02.06. The situation is more complex because of two factors which are the different behaviors of the countries and the non-synchronized multiple waves of the pandemic.

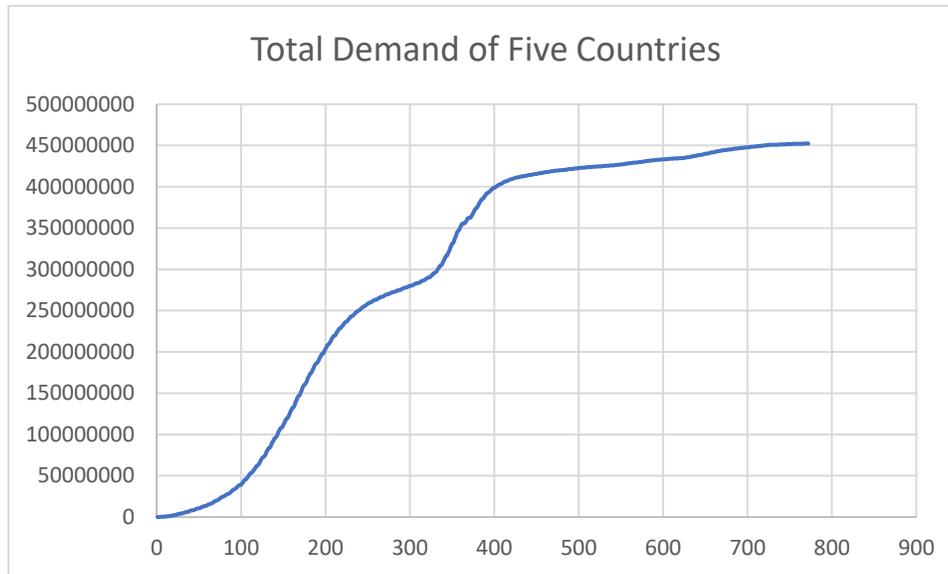

Figure 2. The total demand of Poland, Germany, France, Belgium, and the Czech Republic.

## 3. The problem and basic assumptions

It is assumed that a new pandemic occurs in the future. If it is caused by a known disease, then vaccine may exists immediately. Otherwise, a certain amount of time is needed two develop the new vaccine. The two cases are not distinguished in this paper. Day 1 is the first day when the vaccine can be purchased and the vaccination is started. The vaccination period ends when everyone has received the vaccine. This is a limited period if the supply is continuous and does not converges to 0.

The pandemic occurs in each of a set of countries. The countries have agreed in advance to get the vaccine together. The demand of the vaccine changes in time and is deterministic in each country. The aggregated demand of the group of the countries is the sum of the individual demands.

The vaccine is produced by a single pharmacological company. The production rate is lower than would be necessary to fully satisfy the combined demand continuously. In other words, there are [1,$t$] time periods such that the total production within the period is less than the total aggregated demand in the same period.

The countries claim that the producer must follow an *a priori* determined distribution mechanism. It must be fair and efficient. The main goal of this paper is to determine the algorithm of the mechanism.

The mechanism must balance among conflicting aspects. If a country does not get supply for a longer period, then the pandemic can spread there in a faster way. This country can repeatedly infect the other ones. Thus, no long unsupplied period is allowable in a country. The maximal (relative) shortage should be as low as possible.

The case of dynamic environment is discussed in the second half of the paper. The demands in the countries may change. The forecasted demand can depend on the course of vaccination in the past. The producer may have disruption in the production process.

## 3. The basic model

A discrete time model is presented. The time unit is day, as even the daily production must be huge. There are constraints which are always the same. The application of other constraints depends on the objective function. It is assumed that the production rate is constant, *i.e.* every day the same quantity is produced. The countries may have different importance expressed by weight or population size.

Notations

| | |
|---|---|
| $T$ | The number of the time periods |
| $t$ | The index of the time period (day) |
| $n$ | The number of countries |
| $i$ | Index used for countries |
| $Q$ | The daily production quantity |
| $x_{it}$ | The quantity obtained by country $i$ in time period $t$ |
| $q_{it}$ | The total quantity obtained by country $i$ until time period $t$ |
| $d_i(t)$ | The total demand of country $i$ until the end of time period $t$ |
| $b_{it}$ | Binary variable. It is 1 if there is shortage in country $i$ in time period $t$ |
| $M$ | A big positive number |
| $\varepsilon$ | The tolerated amount of shortage ($\varepsilon > 0$) |
| $h$ | Objective function variable if it is needed |
| $t = 0$ | The index of the initial state |

**Table 1.** Mathematical notations of the models

The notation suggest that the total demands of the countries are provided in the form of a function. It is not necessary. For each country the total demand can be represented by a $T$-dimensional vector.

The initial state is that no country got any vaccine. It is expressed by formulas as follows:

$$\forall i: q_{i0} = x_{i0} = 0 \qquad (1)$$

The daily production must be distributed among the countries.

$$\sum_{i=1}^{n} x_{it} = Q \qquad t = 1,2,\dots,T \qquad (2)$$

The total delivered quantity must be updated in every country and on every day.

$$q_{it} = q_{i,t-1} + x_{it} \quad i = 1,2,\dots,n, t = 1,2,\dots,T \qquad (3)$$

Shortage exists in a country on a day if the total delivered quantity is less than the total demand, *i.e.* $q_{it} < d_i(t)$. This condition is expressed by the inequalities taking into account that the value of $b_{it}$ is either 1, or 0.

$$q_{it} + Mb_{it} \geq d_i(t) \quad i = 1,2,\ldots,n, t = 1,2,\ldots,T \quad (4)$$

If a certain amount of shortage can be tolerated, then the modified version of (4) is as follows:

$$q_{it} + Mb_{it} + \varepsilon \geq d_i(t) \quad i = 1,2,\ldots,n, t = 1,2,\ldots,T \quad (4')$$

The technical constraints describe the nature of the variables.

$$q_{it}, x_{it} \geq 0 \quad i = 1,2,\ldots,n, t = 1,2,\ldots,T \quad (5)$$

and

$$b_{it} = 0 \text{ OR } 1 \quad i = 1,2,\ldots,n, t = 1,2,\ldots,T \quad (6)$$

The simplest version of the model minimizes the number of (country, day) pairs where there is shortage. Thus, the objective function is

$$\min \sum_{i=1}^{n} \sum_{t=1}^{T} b_{it} \quad (7)$$

The countries can be weighted according to their importance by modifying the objective function accordingly.

$$\min \sum_{i=1}^{n} \sum_{t=1}^{T} p_i b_{it} \quad (7')$$

The model is the optimization of the (7) or (7') objective function under the constraints (1), (2), (3), (5), (6), and (4) or (4'). It is what can be considered as efficient variant. A small numerical example is discussed below. It shows this efficient model can put some countries at a disadvantage. An equity model is also required to obtain a proper compromise.

The maximal/maximal weighted shortage is minimized in the equity model. The (country, day) shortage pairs are not counted in the basic version. Thus, the variables $b_{it}$'s and constraints (4), (4'), and (6) are not used. The objective function variable $h$ represents the maximal or maximal weighted shortage. Assume that the magnitude of the error is expressed in the percentage of the population. Variable $h$ is not allowed to be negative. The corresponding condition is

$$h \geq 0 \quad (8)$$

The shortage is $d_i(t) - q_{it}$. If this quantity is negative, then there is no shortage. This case is solved by constraint (8). The weighted shortage is obtained if this quantity is divided by the size of the population. An upper bound of the is obtained by the inequalities

$$h \geq \frac{d_i(t) - q_{it}}{p_i} \quad i = 1,2,\ldots,n, t = 1,2,\ldots,T \quad (9)$$

Thus, the equity model consists of the (1), (2), (3), (5), (8), and (9) constraints and the objective function is

$$\min h \tag{10}$$

If the absolute maximal error is to be minimized, then every $p_i$ in formula (9) must be chosen as 1.

**4. A small numerical example**

The numerical example below is about three fictive countries. However, it simulates a real situation. The basic data of the example can be found in Table 2. The daily production capacity is 1000. The delivered quantities until the end of Day 40 are as follows: Country A 900, Country B: 37,000, Country C: 3100. The cumulative and total demands of the three countries and the total balance in the interval from Day 41 to Day 49 are in Table 2. It is supposed that the daily production quantity, *i.e.* 1000 units, is distributed among the three countries. Column Balance shows that shortage exists from Day 44 to Day 48.

|     | Country |       |      | Total  |         |
|-----|---------|-------|------|--------|---------|
| Day | A       | B     | C    | Demand | Balance |
| 41  | 860     | 37628 | 3174 | 41662  | 338     |
| 42  | 921     | 38575 | 3339 | 42835  | 165     |
| 43  | 985     | 39459 | 3513 | 43957  | 43      |
| 44  | 1054    | 40284 | 3699 | 45037  | -37     |
| 45  | 1127    | 41053 | 3896 | 46076  | -76     |
| 46  | 1205    | 41769 | 4107 | 47081  | -81     |
| 47  | 1289    | 42438 | 4332 | 48059  | -59     |
| 48  | 1379    | 43062 | 4572 | 49013  | -13     |
| 49  | 1476    | 43645 | 4829 | 49950  | 50      |

**Table 2.** The basic data of the numerical example including the demand of the countries, the total demand and the balance.

The optimal solution of the efficient problem without weights and shortage tolerance, *i.e.* the optimization of objective function (7) under the constraints (1), (2), (3), (4) (5), and (6) is as follows:

| DAY | Delivered Quantity | | | Total Delivered Quantity | | | Surplus/Shortage | | |
|-----|----|------|-----|------|-------|------|-----|-----|-----|
|     | A  | B    | C   | A    | B     | C    | A   | B   | C   |
| 41  | 85 | 628  | 287 | 985  | 37628 | 3387 | 125 | 0   | 213 |
| 42  | 0  | 1000 | 0   | 985  | 38628 | 3387 | 64  | 53  | 49  |
| 43  | 0  | 831  | 169 | 985  | 39459 | 3556 | 0   | 0   | 43  |
| 44  | 69 | 788  | 143 | 1054 | 40247 | 3699 | 0   | -37 | 0   |
| 45  | 73 | 730  | 197 | 1127 | 40977 | 3896 | 0   | -76 | 0   |
| 46  | 78 | 711  | 211 | 1205 | 41688 | 4107 | 0   | -81 | 0   |
| 47  | 84 | 691  | 225 | 1289 | 42379 | 4332 | 0   | -59 | 0   |
| 48 | 90 | 670 | 240 | 1379 | 43049 | 4572 | 0 | -13 | 0 |
| 49 | 97 | 646 | 257 | 1476 | 43645 | 4829 | 0 | 0 | 0 |

Table 3. The optimal solution of the efficient problem.

Table 3 shows the drawback of the model. The number of shortage days is 5. It follows from Table 2 that this number cannot be less. However, all shortage days belong to the same country creating a negative discrimination against this country. It is not a serious problem in the example, as the delivered quantities to country B are much higher than the shortage. It means that the vaccination process can be continued without interruption, just not every applicant can get vaccine. However, the efficient approach can cause more serious problems in other cases.

The equity approach, *i.e.* the optimization of objective function (10) under the constraints (1), (2), (3), (5), (8), and (9) gives different results. The solution in Table 4 belongs to the non-weighted case, i.e. $p_1 = p_2 = p_3 = 1$.

| | Delivered Quantity | | | Total Delivered Quantity | | | Surplus/Shortage | | |
|---|---|---|---|---|---|---|---|---|---|
| DAY | A | B | C | A | B | C | A | B | C |
| 41 | 58 | 895 | 47 | 958 | 37895 | 3147 | 98 | 267 | -27 |
| 42 | 0 | 653 | 347 | 958 | 38548 | 3494 | 37 | -27 | 155 |
| 43 | 0 | 884 | 116 | 958 | 39432 | 3610 | -27 | -27 | 97 |
| 44 | 69 | 869 | 62 | 1027 | 40301 | 3672 | -27 | 17 | -27 |
| 45 | 73 | 725 | 202 | 1100 | 41026 | 3874 | -27 | -27 | -22 |
| 46 | 78 | 716 | 206 | 1178 | 41742 | 4080 | -27 | -27 | -27 |
| 47 | 84 | 691 | 225 | 1262 | 42433 | 4305 | -27 | -5 | -27 |
| 48 | 158 | 602 | 240 | 1420 | 43035 | 4545 | 41 | -27 | -27 |
| 49 | 0 | 743 | 257 | 1449 | 43778 | 4802 | -27 | 133 | -27 |

Table 4. The optimal solution of the equity problem. Color red means maximal shortage. Color green is non-maximal shortage.

There are 19 shortage days. 17 out of the 19 shortage days are maximal. Country A might stop the vaccination on days 43 and 49 as there is shortage on these days without delivery. On the other hand, the shortage if any, is always low in every day and country.

The notation $d_i(t)$ suggest that the demand is a function of time. If it is expressed by a formula which can be used in the constraints, then the application of continuous time becomes possible. The constraints get more compact form. However, the problem becomes a hard non-linear problem. Moreover, the discrete time cannot be completely eliminated as the decision on deliveries is made on a daily or weekly basis. If the demands are described by the sequence of these daily/weekly values, then the non-linearity is eliminated from the problem. Instead of the real demand values which become known *a posteori*, the values obtained from the regression can be used.

**5. Application in dynamic environment**

A new pandemic may have different parameters than COVID-19 had and discussed in the *a posteori* analysis of (Pınarbaşı, Vizvári, 2024). On the other hand, the methods of this paper can be applied to uncover the set of papameters. The regression analysis can be done every day. The arcus tangent function can be well approximated. Thus, reasonable forecast can be obtained which can be used in decision making, see (Pınarbaşı, Vizvári, 2024; section 3, page 1033). If the vaccination process is going on, then the length of the proper forecast becomes larger. Thus, it is possible to make daily decisions on the distribution.

The dynamic environment has a computational advantage, too. The optimization problem is updated every day. The past has no variables in the problem. To optimize until the end of the pandemic also has no sense. It is enough to look ahead until the regression seems to be reliable. Thus, the size of the optimization problem is reduced significantly. This makes the problem manageable for a larger group of countries.

To achieve better compromise, it is possible to develop new models using elements from both the efficient and equity models. An easy linear constraint is that an upper bund is introduced to the weighted shortage. If $1 > \delta > 0$ is the allowed maximal relative shortage then the constraints

$$\delta \geq \frac{d_i(t) - q_{it}}{p_i} \qquad i = 1,2,\ldots,n, t = 1,2,\ldots,T$$

must be claimed. Let $s_{it}$ the length of a shortage interval of country $i$ just before day $t$. It is zero if there is no shortage on day $t-1$. The prescribed upper bound of a shortage interval is denoted by $\beta$. Then, the constraints which ensure that there is no longer shortage interval than $\beta$ are as follows:

$$s_{it} + \sum_{l=t}^{t+\beta-s_{it}} b_{it} \leq \beta \qquad i = 1,2,\ldots,n, t = 1,2,\ldots,T.$$

## 6. Conclusion

The paper discusses the problem how to distribute vaccines among cooperative countries. The shortage of vaccine is unavoidable in the case of a new pandemic almost surely. Two mathematical models of fair distribution are presented. One of them is efficient. Its objective is to minimize the number of pairs (*country, day*) such that there is shortage. The other model is based on the concept of equity. It minimizes the maximal weighted shortage. Possible combination of the two models are also discussed. A numerical example shows what can be expected from the models.

Future research must solve larger problems based on the available data of COVID-19. The results obtained in this way are useful in the occurrence of future pandemics. Further combinations of the basic model should also investigated to obtain better compromises.

(database 2023) [The complete Our World in Data COVID-19 dataset - owid-covid-data.csv - ADH Data Portal (africadatahub.org)](#)